%% file: main.tex
\documentclass[conference]{IEEEtran}
\IEEEoverridecommandlockouts
\usepackage{cite}
\usepackage{amsmath,amssymb,amsfonts}
\usepackage{algorithmic}
\usepackage{graphicx}
\usepackage{textcomp}
\usepackage{xcolor}
\usepackage{stfloats}
\usepackage{flushend}

\def\BibTeX{{\rm B\kern-.05em{\sc i\kern-.025em b}\kern-.08em
    T\kern-.1667em\lower.7ex\hbox{E}\kern-.125emX}}

\input{preamble.tex}
\begin{document}

\title{Measuring Multi-Source Redundancy in Factor Graphs 
}

\author{\IEEEauthorblockN{Jesse Milzman, Andre Harrison, Carlos Nieto-Granda, John Rogers}
\IEEEauthorblockA{
\textit{DEVCOM Army Research Laboratory}\\
Adelpi, MD USA \\
\{jesse.m.milzman.civ, andre.v.harrison2.civ, carlos.p.nieto2.civ, john.g.rogers59.civ\}@army.mil}

}

\maketitle

\begin{abstract}
Factor graphs are a ubiquitous tool for multi-source inference in robotics and multi-sensor networks.
They allow for heterogeneous measurements from many sources to be concurrently represented as factors in the state posterior distribution, so that inference can be conducted via sparse graphical methods.
Adding measurements from many sources can supply robustness to state estimation, as seen in distributed pose graph optimization.
However, adding excessive measurements to a factor graph can also quickly degrade their performance as more cycles are added to the graph.
In both situations, the relevant quality is the redundancy of information.
Drawing on recent work in information theory on partial information decomposition (PID),
we articulate two potential definitions of redundancy in factor graphs, both within a common axiomatic framework for redundancy in factor graphs.
This is the first application of PID to factor graphs, and only one of a few presenting quantitative measures of redundancy for them.
\end{abstract}


\section{Introduction}

Autonomous agents operating within an unknown environment must often simultaneously solve several related yet distinct optimization problems.
Concurrently, they model their surrounding environment, estimate their own pose trajectory, the poses of other agents with respect to themselves and the environment, and classify, localize, and track objects within that environment.
These optimization tasks are often modeled in robotics research as a factor graph consisting of variables which represent the unknown quantities and factors that define functions that act on a subset of the variables. 
Typically factors operate only on a limited set of variables, corresponding to measurements on those variables.   

Moreover, when several of these agents operate together information is shared amongst them to generate faster and more accurate estimates of the given target variable(s) that they need to estimate. 
In some of the most austere environments these autonomous systems will form a distributed mobile ad-hoc network (MANET) as there may be no supporting communication infrastructure or equipment.
A recurrent research question in these scenarios is how to maintain or even improve estimation accuracy (and speed) using the least amount of information, in order to both minimize communication between agents as well as reduce the number of variables incorporated into the optimization problem.
This can be done by estimating the individual or sets of information sources that provide the most unique or synergistic information for estimating the target variable while simultaneously minimizing information from redundant sources of information \cite{wang2019factor}.  
While minimizing the communication of redundant information can also reduce computational complexity, it can make the network brittle as the corruption or invalidation of any information source will reduce the estimation accuracy of the target variable(s).
Incorporating a certain amount of redundancy for an information source or set of sources may lead to a more robust distributed estimation network.

Redundancy also plays an important role in the implementation of resilient estimation.
Alternative communication and information-gathering pathways can facilitate adaptation to unexpected failure modes \cite{prorok2021beyond}.
Cummingham \textit{et al.}  proposed one the first distributed implementations of Smoothing and Mapping  (DDF-SAM) resulting in systems with increased resilience to node failures and network topology changes \cite{cunningham2010ddf}.
Khedekar et al. \cite{khedekar2022mimosa} proposed a multi-modal approach to enable resilient SLAM through the use of redundant sets of heterogeneous sensors.

To summarize, in many autonomy scenarios, adding heterogeneous sensors or additional agents can be used to increase the overall robustness and resilience of factor graph-based SLAM estimates.
However, precise measures of the amount of redundancy from difference sources of information present within a factor graph hasn't received as much attention.
Wang \textit{et al.} proposed the use of maximal information content (MIC) to estimate the similarity between different features within a factor graph as part of an unsupervised feature selection approach (FGUFS) \cite{wang2019factor}.

The partitioning of information sources at the level of granularity --- to fully capture the uniqueness, synergy, and redundancy among information sources  --- has not historically been possible with standard information theoretic concepts of mutual information or conditional mutual information ~\cite{wollstadt2021rigorous}.
Recent developments in information theory have seen the advancement of the partial information decomposition (PID) of multivariate information to address such questions.
At its core, PID untangles these notions by hierarchically decomposing the information content among many variables with a redundancy function.
However, this approach, while it has been applied to problems in feature selection and information fusion ~\cite{wollstadt2021rigorous, nikovski2022multi} has not been applied to factor graph analysis.  

In this paper, we present the initial development of a framework for estimating the redundancy within a factor, building upon PID.
We present two measures of redundancy: the direction application of PID as an information-theoretic measure, and a statistical distance-based approach with an analogous form.

\textit{The contributions made in this paper are the following:}

\begin{itemize}
    \item Building upon the work in PID, we outline a formal axiomatic framework for defining multi-source redundancy in factor graphs.
    \item We define two potential redundancies within this framework: an information-theoretic redundancy and a statistical distance-based redundancy.
    \item We investigate these definitions analytically for supplemental factors in linear factor graphs, rewriting them as closed-form integrals on the base graph.
    \item We apply these to simulations of a 2D landmark SLAM task, in order to compute the redundancy of the information contributed by two landmarks.
\end{itemize}


\section{Preliminaries}
\label{section:preliminaries}
In this section, we briefly review the key definitions from the PID framework, as well as those for linear factor graphs.
For factor graphs in particular, we will largely use our own notation in order to simplify the later presentation of our PID-based redundancy framework in Sec.~\ref{sec:redundancyLFGs}.
This work is primarily focused on linear factor graphs.

\subsection{Partial Information Decomposition}
\label{subsection:PID}

The partial information decomposition was originally put forward in \cite{williams2010nonnegative} as a one-sided multivariate extension of mutual information, and as a resolution to the signed nature of interaction information.
In its most common form, PID decomposes the information $I(T; X_1, ..., X_n)$ into $D(n)-2$ components, where $D(n)$ is the $n$-th Dedekind number.
For two sources $X_1$ and $X_2$, this resolves into the so-called bivariate PID, i.e. the sum $I(T;X_1,X_2) = S + R + U_1 + U_2$, with each summand (atom) respectively referred to as the synergistic, redundant, and unique information(s).
Crucially, every PID is uniquely defined by a redundancy function $\Ired$, such as the $\Imin$ function defined in \cite{williams2010nonnegative}.
In the bivariate case, the redundancy function determines $R$.
More generally, it determines the redundant information shared between multiple sources, formalized as collections of random variables.

The PID has primarily been applied to theoretical and experimental neuroscience \cite{EntropySIOverview2018}, although there have been a few works applying its tools to other areas \cite{chan2017gene}.
Although the framework was received with great interest from interdisciplinary information theorists and complex systems scientists, PID remains ill-posed due to the multiplicity of proposed redundancy functions.
The original $\Imin$ redundancy function remains controversial, since it does not distinguish between sources providing the same information regarding a target outcome from the same \textit{amount} of information \cite{EntropySIOverview2018}.
At present, we do not find this critique relevant to the application at hand, as we are more concerned with choosing a posterior that sufficiently constrains the target variables for accurate state inference.

The $\Imin$ redundancy function takes the PID lattice of source antichains as its domain, so we must first define that set.
Let $Z_1, ..., Z_n \in \mathbb{R}^N$ be a collection of random 
variables in our $N$-dimensional configuration space, referred to as predictor variables, and $\bm{Z} = (Z_i)_{i=1}^N$ be the full $n \times N$ vector.
In keeping with many previous PID works including \cite{williams2010nonnegative}, when it will not cause confusion, we will identify sets of indexed variables with vectors ordered along increasing index, i.e. $\bm{Z} = (Z_i)_i = \{ Z_i \}_i$.
Let $\mathcal{P}(A)$ to denote the powerset of any set $A$, and $[z] = \{ 1, ..., z\}$ for any integer $z$.
Then $\mathcal{P}(\bm{Z})$ denotes the collection of \textbf{sources}, i.e. subsets of predictor variables, which we denote $\bm{Z}_{J}$ for some index vector $J \in \mathcal{P}([n])$.
The PID lattice, denoted $\mathcal{A} (\bm{Z})$, is the collection of antichains of the poset $(\mathcal{P}(\bm{Z}), \subseteq)$.
Put differently, the PID lattice is comprised of all possible collections of sources in $\mathcal{P}(\bm{Z})$ such that no two sources in the same collection are comparable by inclusion. Put formally, for any nonempty $\alpha \in \mathcal{P}(\mathcal{P}(\bm{Z}))$,
\begin{equation}
\label{eq:defn:antichain}
\alpha \in \mathcal{A}(\bm{Z}) \Leftrightarrow \forall \bm{Z}_J,\bm{Z}_{J'} \in \alpha, \, \bm{Z}_J \not\subset \bm{Z}_{J'}.
\end{equation}
The antichain lattice $\mathcal{A}(\bm{Z})$ comes equipped with the following partial ordering:
\begin{equation}
\label{eq:defn:antichain.order}
\alpha \preceq \beta \Leftrightarrow \forall \bm{Z}_{J} \in \beta, \exists \bm{Z}_{J'} \in \alpha, \, J' \subseteq J
\end{equation}
Thus, as one ascends the lattice, individual sources within the antichain will encompass more variables, non-monotonically decreasing the number of distinct sources in the antichain by the top. In particularly, if $\alpha_{\min} = \{ \{ \bm{Z}_i \} \}_i$ (i.e. the antichain of singletons) and $\alpha_{\max} = \{ \bm{Z} \}$ (i.e. the source containing all predictors), we have $\alpha_{\min} \preceq \beta \preceq \alpha_{\max}$ for all $\beta \in \mathcal{A}(\bm{Z})$.
When convenient, we may use an abuse of notation whereby the antichain $\alpha$ contains the index sets for its sources, e.g. $J \in \alpha$.
In this work, every source variable $Z_i$ has a smooth density $p_{Z_i}$ supported everywhere on the configuration space. Moreover, every joint density $p_{\bm{Z}_{J}}$, including $p_{\bm{Z}}$, is smooth and supported everywhere on the product space.

Let the target variable $T \in \mathbb{R}^M$ be given, and assume that it has a smooth joint density $p_{\bm{Z},T}$ with the predictors, supported everywhere.\footnote{We assume full support to simplify our presentation.}
The specific information of a predictor source $\bm{Z}_{J}$ toward the target variable is the integral function defined pointwise on target realizations $T=t$.
\begin{definition}[Specific Information]
\label{defn:specificInfo}
Let the predictors and target variable and $(\bm{Z}, T)$ have a smooth joint density $p_{\bm{Z},T}$.
Then for any $t$ such that $p_{T}(t) > 0$, we define the specific information of $\bm{Z}_J$ about $T=t$ to be
\begin{align}
    I_{\bm{Z}_j}(t)
     &= D \left( \, p_{\bm{Z}_J|T}(\bm{Z}_J|t) \; || \; p_{\bm{Z}_J}(\bm{Z}_J) \right).
\end{align}
Moreover, we let $I_{\bm{Z}}(t) = 0$ whenever $p_{T}(t) = 0$.
\end{definition}
We emphasize that we use the subscript $\bm{Z}_J$ to label our specific informations, but they are not functions of the variable $\bm{Z}_J$, only $T$.
The $\Imin$ redundancy function is defined on $\mathcal{A}(\bm{Z})$ as follows:
\begin{definition}[$\Imin$ Redundancy Function]
    \label{defn:Imin}
    Let $\alpha \in \mathcal{A}(\bm{Z})$ be the antichain composed of sources $\alpha = (\bm{Z}_{J_k})_k$.
    Then
    \begin{align}
        \Imin(\alpha;T) &= \mathbb{E} \min_{k} I_{\bm{Z}_{J_k}}(T)
    \end{align}
\end{definition}
\begin{remark}
    The $\Imin$ redundancy function satisfies several desirable properties:
    \begin{itemize}
        \item \textbf{Nonnegativity (P)}
        \begin{equation}
        \label{eq:PIDaxiom.P}
        \Imin(\alpha;T) \geq 0 \quad \forall \alpha \in \mathcal{A}(\bm{Z})
        \end{equation}
        \item \textbf{Monotonicity (M)}
        \begin{equation}
        \label{eq:PIDaxiom.M}
            \Imin(\alpha;T) \leq \Imin(\beta;T) \quad \text{ if }\beta \subseteq \alpha
        \end{equation}
        \label{eq:PIDaxiom.SR}
        \item \textbf{Self-redundancy (SR)}
        \begin{equation}
            \Imin(\{\bm{Z}_J\}, T) = I(\bm{Z}_J, T)
        \end{equation}
    \end{itemize}
    The latter two of these are typically enumerated among the Williams-Beer (WB) axioms, along with a symmetry property that follows from $\Imin$ being well-defined on $\mathcal{A}(\bm{Z})$.
\end{remark}

This structure allows the $\Imin$ function to be used to decompose the full information $I(T; \bm{Z})$ into nonnegative atoms in the discrete case, using the M\"{o}bius inversion of $\Imin$, typically denoted $\Pi$:
\begin{equation}
    \Pi(\alpha) = \Imin(\alpha) - \sum_{\beta \prec \alpha} \Imin(\beta)
\end{equation}
In the bivariate case, the redundancy, synergy, and unique information are given by $R= \Pi(\{Z_1\}\{Z_2\}), S = \Pi(\{\bm{Z}\})$, and $U_i = \Pi(\{\bm{Z}_i\})$.
Although these other atoms may be of future interest for the analysis of factor graphs, in this work we restrict our attention to redundancies.

\subsection{Linear Factor Graphs}
\label{subsection:LFGs}
Factor graphs~\cite{kschischang2001factor} are graphical models that are well suited to modeling complex perception estimation problems in robotics, such as Simultaneous Localization and Mapping (SLAM), or Structure from Motion (SFM)~\cite{dellaert2017factor}~\cite{dellaert2021factor}. 
Similar to other probabilistic graphical models (PGMs) such as Bayesian networks and Markov random fields, factor graphs allow high-dimensional state variables to be decomposed into components with sparse dependencies.
A common application of factor graphs in robotics is in pose graph optimization (PGO), in which there have been many advancements in recent years.
For centralized PGO, there has been a development of certifiably correct estimation methods that are capable of efficiently recovering globally optimal solutions of generally intractable estimation problems~\cite{bandeiraCRM2016}~\cite{rosen19ijrr}. 
In distributed PGO, Tian et al.~\cite{Tian2021Distributed} presented the first Distributed Certifiably Correct Pose- Graph Optimization (DC2-PGO).

Given all measurements and priors on the state $\bm{z}$, factor graphs formally factor the posterior density $p(\bm{x}|\bm{z})$ into the product of functions specific to each measurement, typically unnormalized densities of the relationship between a state observable $h_i(\bm{x})$ and the corresponding measurement $z_i$.
For instance, in SLAM and PGO, measurements of one or two keyframe(s) from the trajectory are pulled from odometry, IMU, LiDAR scans, and/or visual perception.
Taken together, all these measurements allow for the inference of the full trajectory from its posterior.
However, by representing this distribution in a factor graph, the maximal a posteriori (MAP) trajectory estimate can be computed piecemeal by exploiting the sparsity of the graph~\cite{dellaertIJRRsqrtsam}.
For the theoretical elements of this paper, we will use the closed-form expression for the MAP of a linear factor graph.

Consider a system comprised of $n$ variables in $N$-dimen-\\sional configuration space, i.e. $X_1, ..., X_n \in \mathbb{R}^N$. 
A factor graph is comprised of these variables and an associated collection of factors $\mathcal{F} = \{ \phi_{j} \}_{j=1}^m$, which is a collection of real-valued functions, with each $\phi_j$ taking a subset of variables $\bm{X}_{\bm{i}_j}$ as its argument.
Most often, factors take one or two variables as their argument.
Formally, a factor graph is the bipartite graph $\mathcal{G} = (\mathcal{V},\mathcal{E})$, where $\mathcal{V} = \mathcal{F} \cup \bm{X}$ and $\{ i,j \} \in \mathcal{E}$ iff $i \in \bm{i}_j$, i.e. iff $X_i$ is an argument for $\phi_j$.
In robotic state estimation, univariate factors can encode priors (including starting position and exogenous map information) and GPS measurements, while bivariate factors might encode wheel odometry, visual perception, or LiDAR point-cloud comparisons \cite{choudhary2017distributed}.
In this work, we model every factor as a multivariate Gaussian distribution, produced by a perfectly calibrated Gaussian measurement.
We use $\mathcal{N}_{z}(m, \Sigma)$ to represent Gaussian densities at $Z=z$ for mean $m$ and covariance matrix $\Sigma$.
For any vector $z$, we denote the Mahalanobis norm with its precision matrix, i.e. $\left\| z \right\|^2_M = z^\top M z$, since we will more often work with precision matrices instead of noise matrices.
Every factor $\phi_j$ is given by:
\begin{equation}
\label{eq:Gaussian_factor}
\phi_j (\bm{x}_{\bm{i}_j}) = \exp -\frac{1}{2} \left\|  
\bm{A}_j \bm{x} - z_j  \right\|_{\Gamma_j}^2
\end{equation}
for some $N \times n N$ measurement matrix $\bm{A}_j$.
This is analogous to the Jacobian matrix of the measurement operator in nonlinear factor graphs \cite{dellaert2017factor}.

We will often merely refer to the collection of factors $\mathcal{F}$ as the factor graph, and subsets of it as subgraphs.
Such subsets are denoted $\mathcal{F}_{J} \subseteq \mathcal{F}$ for some vector of factor indices $J \subset [m]$.
Using powerset notation, the collection of all subgraphs forms a partially ordered set (poset) under inclusion: $(\mathcal{P}(\mathcal{F}), \subseteq)$.
Note that $\mathcal{F}_{J'} \subset \mathcal{F}_{J}$ iff $J' \subset J$, and we will often prefer the latter notation for convenience.
Any factor subgraph $\mathcal{F}_{J}$ is associated to the product factor $\Phi_{\mathcal{F}_{J}} = \prod_{j \in J} \phi_j(\bm{x}_{\bm{i}_j})$.
By concatenating the measurements and measurement matrices, we may rewrite these
\begin{equation}
    \Phi_{\mathcal{F}_{J}} (\bm{x}) = \exp -\frac{1}{2} \left\| \bm{A}_{J} \bm{x} - \bm{z}_{J}  \right\|_{\Gamma_{J}}^2 \\
\end{equation}
where
\begin{align}
    \nonumber A_{J} = \begin{bmatrix} \bm{A}_{j_1} \\ \vdots \\ \bm{A}_{j_{|J|}}
    \end{bmatrix},
    & \quad
    \bm{z}_{J} = \begin{bmatrix} \bm{z}_{j_1} \\ \vdots \\ \bm{z}_{j_{|J|}} \end{bmatrix}, \quad
    \nonumber \Gamma_{J} =
    \begin{bmatrix}
        \Gamma_{j_1} & \bm{0} & \hdots \\
        & \ddots & \\
        \hdots & \bm{0} & \Gamma_{j_{|J|}}
    \end{bmatrix}
\end{align}
For the full graph $\mathcal{F}$, we omit subscripts on $\bm{A}$, $\bm{z}$, and $\mathcal{F}$.
In the event that $\rank(\bm{A}_{J})=n$, the product factor $\Phi_{\mathcal{F}_{J}}$ presents the unnormalized density for a joint distribution on $\bm{X}$.
In this situation, we say that $\mathcal{F}_{J}$ is full-rank.
By solving the quadratic form in the exponential, i.e. refactoring it into the standard Gaussian form, we have
\begin{align}
\label{eq:factorProductFunction.densityForm}
    \Phi_{\mathcal{F}_{J}} ( \xx ) & \propto \mathcal{N}_{\xx} (\bm{\mu}_{J}, \Lambda_{J}^{-1}) \\
\label{eq:factorProductFunction.densityForm.lambda} \Lambda_{J} &= \bm A_{J}^\top \Gamma_{J} \bm A_{J}\\
    \label{eq:factorProductFunction.densityForm.mu} \bm{\mu}_{J}  &= \Lambda_{J}^{-1} \bm A_{J}^\top \Gamma_{J} \bb
\end{align}
where, as before, we omit subscripts when referencing the full graph $\mathcal{F}$.
In fact, for any full-rank $\mathcal{F}_{J}$, this distribution $\Phi_{\mathcal{F}_{J}}$ corresponds to posterior state distribution after accounting for the included measurements, i.e.
\begin{equation}
    p(\bm{x} | \bm{z}_{J}) \propto \Phi_{\mathcal{F}_{J}}(\bm{x})
\end{equation}
Viewed from this perspective, we see that $\mu_{J}$ is the maximum \textit{a posteriori} (MAP) state estimate \cite{dellaert2017factor,dellaert2021factor}.

In practice, the dimensionality of $\Lambda$ will be quite large, and it is often computationally difficult if not intractable to naively solve for $\bm \mu$.
Factor graphs are employed in contexts in which we expect them to be sparse, i.e. we expect $\bm{A}$ to be sparse, or (as the Jacobian) sparse at every point for nonlinear factor graphs.
This allows one to exploit this sparsity to solve for $\bm{\mu}$ iteratively \cite{dellaert2017factor}. 

We assume that every measurement $z_j$ is independently sampled, each from the conditional distribution $p(z_j | \bm{x}) \sim \mathcal{N}(\bm{A}_J \bm{x}, \Gamma_{z_j}^{-1})$.
Note that the density of this distribution is merely a normalization of $\phi_j$.
This assumption will allow us to estimate the information content of individual factors, given some prior on $\bm{X}$.
However, in practice, all of our information regarding $\bm{X}$ is encoded in $\mathcal{F}$.
Rather than assuming a prior distribution separate from $\mathcal{F}$, we instead partition our factors into two sets: base factors (forming the base graph) and supplementary factors.
$$
\mathcal{F} = \mathcal{F}_{B} \cup \mathcal{F}_{\bar{B}}
$$
where $\bar{B} = [m] \setminus B$.
Taking the density induced by the base graph as our prior distribution, we denote our prior and posteriors as:
\begin{align}
\label{eq:pxPrior}
p_{\bm{X}}(\bm{x}) &:= p(\bm{x} | \bm{z}_{B}) \propto \Phi_{\mathcal{F}_{B}}\\
\label{eq:pxPosterior}
p_{\bm{X}|\bm{Z}_J}(\bm{x} | \bm{z}_{J}) &:= p(\bm{x} | \bm{z}_{B \cup J}) \propto \Phi_{\mathcal{F}_{B \cup J}}
\end{align}
where we omit subscripts $p(\bm{x}), \, p(\bm{x}|\bm{z}_J)$ where they are redundant.
To simplify notation, we denote $\tilde{J} = B \cup J$ for all $J \subset \bar{B}$, so that we have $p(\bm{x}|\bm{z}_{J}) \sim \mathcal{N}_{\bm{x}}(\bm{\mu}_{\tilde{J}}, \Lambda_{\tilde J})$ as per (\ref{eq:factorProductFunction.densityForm}).
Moreover, from here on, we will employ $\Delta_{J}$ in place of $\Lambda_{J}$ for the matrix expression in (\ref{eq:factorProductFunction.densityForm.lambda}) when it is ambiguous as to whether $\mathcal{F}_{J}$ is full-rank, i.e. whether $\Delta_{J}$ is nonsingular.
If $\Delta_J$ is singular, then $\bm \mu_J$ is not defined.
Even if $\Delta_J$ is singular, it will still be symmetric and positive semi-definite, and thus $\mathcal{F}_{\tilde J}$ will be full-rank, as we will see below (\ref{eq:prop.union_lambda.tildeJ}).
Hence, (\ref{eq:pxPosterior}) and $\bm{\mu}_{\tilde J}$ are always well-defined.

We summarize this exposition by introducing the shorthand of a supplemented factor graph, i.e. factor graphs with such a partitioning of their factors as described above.
\begin{definition}[Supplemented Factor Graph]
\label{defn:SFG}
    A supplemented factor graph is represented by the triplet $(\bm{X}, \mathcal{F}, B)$, where $\bm{X}$ are the target variables, $\mathcal{F}$ is a collection of $m$ linear Gaussian factors of the form (\ref{eq:Gaussian_factor}), and $B \subset [m]$ is a subset of indices such that $\mathcal{F}_B$ is full-rank for $\bm{X}$. The elements of $\mathcal{F}_B$ are referred to as base factors, and determine the prior distribution $p_{\bm{X}}$ as per (\ref{eq:pxPrior}). We refer to $\mathcal{F}_{\bar B}$ as the supplemental factors, $\bm{Z}_{\bar{B}}$ as the supplemental measurements, and any $\mathcal{F}_{\tilde J}$ as a supplemented subgraph of $\mathcal{F}$. A supplemented subgraph $\mathcal{F}_{\tilde J}$ determines the $\bm{Z}_{\tilde J}$-posterior distribution $p_{\bm{X}|\bm{Z}_J}$ as per (\ref{eq:pxPosterior}).
\end{definition}

Now that we have meaningful prior and posterior distributions, we will be able to quantify and decompose the information provided by the supplementary factors in this model.
To conclude this section, we provide some useful properties that we will need in the next section.

\begin{proposition}
Let $J,K \subset [m]$ be mutually exclusive sets of factors.
Then we have that:
\begin{align}
\label{eq:prop.union_lambda} \Delta_{J \cup K} &= \Delta_J + \Delta_K \\
\intertext{If $\mathcal{F}_{J \cup K}$ is full-rank,}
\label{eq:prop.union_mu} \bm{\mu}_{J \cup K} &= \Lambda_{J \cup K}^{-1} \left( \bm{A}_{J}^\top \Gamma_{J} \bm{z}_{J} + \bm{A}_{K}^\top \Gamma_{K} \bm{z}_{K}  \right)
\end{align}
In particular, for a given base graph $B$ and supplementary set $J \subset \bar{B}$, we have that
\begin{align}
\label{eq:prop.union_lambda.tildeJ}
\Lambda_{\tilde J} &= \Lambda_{B} + \Delta_{J} \\
\label{eq:prop.union_mu.tildeJ}
\bm{\mu}_{\tilde J} &= \Lambda_{\tilde J}^{-1} \left( \Lambda_B \bm{\mu}_{B} + \bm{A}_{J}^\top \Gamma_{J} \bm{z}_J \right) 
\end{align}
\end{proposition}
\begin{proof}
By expanding $\bm{A}_{J\cup K}, \bm{z}_{J \cup K},$ and $\Gamma_{J \cup K}$ into their $J$ and $K$ block components, we arrive at (\ref{eq:prop.union_lambda}) by direct computation of (\ref{eq:factorProductFunction.densityForm.lambda}).
Similarly, (\ref{eq:prop.union_mu}) follows from (\ref{eq:factorProductFunction.densityForm.mu}).
\end{proof}
Moreover, the poset of subgraphs is compatible with the Loewner ordering on their information matrices.
\begin{proposition}
\label{prop:Loewner.subsource}
For any $J,J' \subseteq [m]$,
\begin{equation}
J' \subseteq J \Rightarrow \Delta_{J'} \leq \Delta_J
\end{equation}
\end{proposition}
\begin{proof}
Follows readily from (\ref{eq:prop.union_lambda}).
\end{proof}

Using the prior/posterior from (\ref{eq:pxPrior}-\ref{eq:pxPosterior}), we may compute the mutual information for any supplemented factor graph $\mathcal{F}_{\tilde{J}} \subset \mathcal{F}$.
\begin{proposition}
    \label{prop:MIforLFGs}
    Let $\mathcal{F}_{\tilde{J}} \subset \mathcal{F}$ be a supplemented subgraph for some $J \subset \bar{B}$.
    Then the mutual information between factor measurements $\bm{Z}_J$ and the state $\bm{X}$ are given by
    \begin{equation}
        I(\bm{Z}_J ; \bm{X})
        = \frac{1}{2} \log \frac{|\Lambda_{\tilde J}|}{|\Lambda_{B}|}
        = \frac{1}{2} \log |I + \Delta_J \Lambda_B^{-1}|
    \end{equation}
\end{proposition}
\begin{proof}
We recall that for a Gaussian vector $U \sim \mathcal{N}(\mu, \Sigma)$, the differential entropy is given by
$h(U) = \frac{1}{2} \ln |2 \pi e \Sigma|$ \cite{cover2006elements}.
Thus, using (\ref{eq:pxPrior}-\ref{eq:pxPosterior}) and (\ref{eq:factorProductFunction.densityForm}-\ref{eq:factorProductFunction.densityForm.lambda}), we compute
\begin{align}
    \nonumber I(\bm{Z}_J ; \bm{X}) &= h(\bm{X}) - h(\bm{X} | \bm{Z}_J)\\
    \nonumber &= \frac{1}{2} \log \frac{|2 \pi e \Lambda_B^{-1}|}{|2 \pi e \Lambda_{\tilde J}^{-1}|}
    = \frac{1}{2} \log \frac{|\Lambda_{\tilde J}|}{|\Lambda_{B}|} 
\end{align}
\end{proof}
This computation gives us an information-theoretic monotonicity for the poset of subgraphs.
\begin{corollary}
\label{corollary:MIFG_monotonicity}
    For any pair of full-rank subgraphs $\mathcal{F}_J, \mathcal{F}_{J'} \subseteq \mathcal{F}$,
    \begin{equation}
    J' \subseteq J \Rightarrow I(\bm{Z}_{J'} ; \bm{X}) \leq I(\bm{Z}_J ; \bm{X}).
    \end{equation}
\end{corollary}

There are a few conditional moments of $\bm{\mu}_{\tilde J}$ that we will need in the next section.
We enumerate those here:
\begin{proposition}
\label{prop:muTildeJ.moments}
    Let supplemented subgraph $\mathcal{F}_{\tilde J} \subset \mathcal{F}$ be given, with prior and posterior distributions as in (\ref{eq:pxPrior}-\ref{eq:pxPosterior}).
    Then we have the following conditional moments for all matrices $\bm T$ and vectors $\bm{m}$:
    \begin{align}
    \label{eq:muTildeJ.firstMoment}
    \mathbb{E} \left( \bm{\mu}_{\tilde J} \; \bigg| \;  \bm{x} \right)
    &= \Lambda_{\tilde J}^{-1} \left( \Lambda_B \bm{\mu}_B + \Delta_J \bm{x} \right) \\
    \nonumber
    \mathbb{E} \left( \left\| \bm{\mu}_{\tilde J} + \bm{m} \right\|_{\bm{T}}^2 \; \bigg| \; \bm{x} \right)
    &= \tr \left( \bm{T} \Lambda_{\tilde J}^{-1} \Delta_J \Lambda_{\tilde J}^{-1} \right)\\
    \label{eq:muTildeJ.secondMoment.general}
    & + \left\| \Lambda_{\tilde J}^{-1} \left( \Lambda_B \bm{\mu}_B + \Delta_J \bm{x} \right) + \bm{m} \right\|_{\bm{T}}^2
\end{align}

\end{proposition}
\begin{proof}
    The first moment (\ref{eq:muTildeJ.firstMoment}) follows readily from (\ref{eq:prop.union_lambda.tildeJ}) and $\mathbb{E} \left( \bm{Z}_J \mid \bm{x} \right)= \bm{A}_J \bm{x}$.
    Then let $\bm{m}' = \mathbb{E} ( \bm{\mu}_{\tilde J} \mid \bm{x} )$ be this conditional mean.
    We rewrite the quadratic form:
    \begin{align}
    \nonumber  \left\| \bm{\mu}_{\tilde J} - \bm{m}' \right\|_{\bm{T}}^2
    &= \left\| \Lambda_{\tilde J}^{-1} \bm{A}_{J}^\top \Gamma_{J} \left( \bm{z}_J - \bm{A}_J \bm{x} \right) 
     \right\|_{\bm{T}}^2 \\
     &= \left\|  \bm{z}_J - \bm{A}_J \bm{x}
     \right\|_{\bm{H}^\top \bm{T} \bm{H}}^2
    \end{align}
    letting $\bm{H} = \Lambda_{\tilde J}^{-1} \bm{A}_{J}^\top \Gamma_{J}$.
    Since $p(\bm{z}_J | \bm{x}) \sim \mathcal{N}(\bm{A}_J \bm{x}, \Gamma_J^{-1})$, in expectation we have that:
    \begin{align}
    \nonumber \mathbb{E}_{p(\bm{z}_J | \bm{x})} \; \left\| \bm{\mu}_{\tilde J} - \bm{m}' \right\|_{\bm{T}}^2
    &= \tr \left(\Gamma_J^{-1} \bm{H}^\top \bm{T} \bm{H} \right) \\
    &= \tr \left( \bm{T} \Lambda_{\tilde J}^{-1} \Delta_J \Lambda_{\tilde J}^{-1} \right)
    \end{align}
    taking advantage of the cyclic property to rearrange the trace.
    The extension to Eq.~(\ref{eq:muTildeJ.secondMoment.general}) is an application of a well-known recentering formula \cite[Sec.~8.2.2]{petersen2008matrix}, whereby for $\bm{U} \sim \mathcal{N}(\bm{v}, \Sigma)$:
    \begin{align}
    \mathbb{E} \left\| \bm{U} - \bm{v}' \right\|^2_{\bm{M}} &= \tr(\bm{M} \Sigma) + \left\| \bm{v} - \bm{v}' \right\|_{\bm{M}}^2.
    \end{align}
\end{proof}

\section{Measuring Redundancy in Linear Factor Graphs}
\label{sec:redundancyLFGs}

In this section, we introduce two metrics for measuring the redundancy of information provided by supplemental factors in a factor graph.
One of these metrics is a straight-forward application of the $\Imin$ redundancy function from PID, while the latter will consider redundancy pointwise on the target variable(s) in a manner analogous to $\Imin$ but from the perspective of statistical distance.
In order to maintain notational similarity between our metrics without causing confusion with PID functions from the literature (which, ultimately, are all extensions of mutual information and denoted with an annotated $I$), we denote the $\Imin$-based metric as $\Rwb$.
We then define another redundancy metric based upon pointwise statistical distance, denoted $\Rw$ in reference to the Wasserstein-2 distance.
These two functions $\Rwb$ and $\Rw$ are by no means the only possibilities, but demonstrate distinct ways to quantify redundancy ``pointwise'' with respect to target variables.

Before we define each metric in turn, we will first introduce a minimal formalism for redundancy that both will follow, generalizing the PID lattice from ~\cite{williams2010nonnegative} to more arbitrary quality metrics of posterior distributions.
The motivation here is to provide a unified language for redundancy, independent of the quality function that is being redundantly guaranteed.

\begin{definition}[Redundancy and Quality Functions]
\label{defn:RQfunctions}
    Let the supplemented factor graph $(\bm{X}, \mathcal{F}, B)$ be given as in Def.~\ref{defn:SFG}.
    Further, consider the measurements $\bm{Z}$ as the collection of predictor variables for target $\bm{X}$ as in Sec.~\ref{subsection:PID}, with the collection of antichains $\mathcal{A}(\bm{Z})$ defined as in Eqs.~(\ref{eq:defn:antichain}-\ref{eq:defn:antichain.order}).
    A \textbf{PID-like redundancy measure} is the pair $(\mathtt{R}, \mathtt{Q})$, where $\mathtt{Q}: \mathcal{P}(\bm{Z}_{\bar B}) \to [Q_{\min}, Q_{\max}]$ is the \textbf{quality function} and $\mathtt{R}: \mathcal{A}(\bm{Z}_{\tilde B}) \to [Q_{\min}, Q_{\max}]$ is the \textbf{redundancy function}, when they fulfill the following axioms:
    \begin{enumerate}
        \item[(MR)]\label{axiom:MR} \textbf{Monotonicity of $\mathtt{R}$} For all $\alpha, \beta \in \mathcal{A}(\bm{Z})$, we have that $\mathtt{R}(\alpha) \leq \mathtt{R}(\beta)$ whenever $\beta \subset \alpha$.
        \item[(MQ)] \textbf{Monotonicity of $\mathtt{Q}$} For all $\bm{Z}_{J}, \bm{Z}_{J'} \in \mathcal{P}(\bm{Z})$, we have that $\mathtt{Q}(\bm{Z}_J) \leq \mathtt{Q}(\bm{Z}_{J'})$ whenever $\bm{Z}_{J} \subseteq \bm{Z}_{J'}$
        \item[(SR)] \textbf{Self-Redundancy} $\mathtt{R}(\{ \bm{Z}_J \}) = \mathtt{Q}(\bm{Z}_J)$ for all $J \subset \bar{B}$.
    \end{enumerate}
\end{definition}
A notable feature of the $\Imin$ PID is that it quantifies redundancy by looking at the minimal specific information provided by any source, in expectation, about the target variable pointwise.
That is, for any realization $X=x$ of the target, $\Imin$ accounts for the least specific information that any source might provide under conditional expectation.
Although this definition of redundancy has generated controversy in the PID literature, we consider it intuitively aligned with what we would want from a redundancy metric in factor graphs.
Namely, redundancy ought to (a) provide robustness guarantees on the minimal information known if one or more sources fail, and (b) identify superfluous information sources in resource-constrained scenarios.
We want this same target-pointwise specificity for any redundancy function.
Thus, we offer the following definition as a further specification of Def.~\ref{defn:RQfunctions}:
\begin{definition}[Target Integral Redundancy and Specific Quality]
\label{defn:targetIntegral}
A PID-like redundancy measure $(\mathtt{R},\mathtt{Q})$ is said to be a \textbf{target integral} if there exists a mapping $\mathtt{S}: \mathcal{P}(\bm{Z}_{\bar B}) \to L^1(\mathbb{R}^{nN}, p(\bm{x}) d\bm{x}),$ denoted $\bm{Z}_J \mapsto \mathtt{S}_{\bm{Z}_J}$, such that
\begin{align}
    \label{eq:defn.targetIntegral.redundancy}
    \mathtt{R}(\alpha) &= \int p(\bm x) \min_{J \in \alpha} \mathtt{S}_{\bm{Z}_J} (\bm x) \, d \bm x \\
    \label{eq:defn.targetIntegral.quality}
    \mathtt{Q}(\bm{Z}_J) &= \int p(\bm x) \, \mathtt{S}_{\bm{Z}_J} (\bm x) \, d \bm x
\end{align}
We refer to this mapping as the \textbf{specific quality} function, with $\mathtt{S}_{\bm{Z}_J}(\bm{x})$ as the specific quality about $\bm{x}$ provided by $\bm{Z}_J$.
\end{definition}
This notion of specific quality is motivated by specific information from \cite{deweese1999measure}, which was used to define PID in \cite{williams2010nonnegative}.
Both $\Rwb$ and $\Rw$ are target integrals, and with Defs.~\ref{defn:RQfunctions}-\ref{defn:targetIntegral} can now be entirely defined by their specific quality functions.
Moreover, it is evident that any target integral redundancy measure will satisfy axioms (MR) and (SR).
Note that computing (\ref{eq:defn.targetIntegral.redundancy}) explicitly requires one to partition the target state space into regions where a given specific quality is minimal.
In practice, it will be easier to numerically approximate this integral with Monte Carlo sampling from $p(\bm{x})$. 

\subsection{An Information-Theoretic Redundancy Measure}

The first redundancy metric we consider is the application of the $\Imin$ PID from \cite{williams2010nonnegative}, (Sec.~\ref{subsection:PID}) to the supplementary measurements in factor graphs (Sec.~\ref{subsection:LFGs}).
Here, the posterior quality metric will be standard mutual information, i.e. $\Qwb \equiv I(\cdot; \bm{X})$, decomposed by the $\Rwb \equiv \Imin( \cdot ; \bm{X})$ redundancy function from Def.~\ref{defn:Imin}.
It is already well-known that $\Rwb$, as the $\Imin$ PID, satisfies (MR) and (SR) axioms in Def.~\ref{defn:RQfunctions}, while (MQ) follows from Cor.~\ref{corollary:MIFG_monotonicity}.

\begin{definition}[Specific Information for LFGs]
Let the supplemented factor graph $(\bm{X}, \mathcal{F}, B)$ be given.
For every supplemental source $\bm{Z}_J \subset\bm{Z}_{\bar B}$, the specific information provided about $\bm{X}=\bm{x}$ is given by, 
\begin{align}
\label{eq:defn.specificQuality.PID}
\Swb_{\bm{Z}_{J}}(\bm{x})
&= \mathbb{E} \left( \log \frac{p(\bm{x} | \bm{Z}_{J})}{p(\bm{x})} \; \bigg| \; \bm{x}  \right)
\end{align}
\end{definition}

For any $(\bm{X},\mathcal{F}, B)$, we may compute the closed-form of this function.
\begin{proposition}
\label{prop.specificInfo}
    \begin{align}
    \label{eq:prop.specificInfo}
        \Swb_{\bm{Z}_{J}}(\bm{x})
        &=
         I(\bm{Z}_J ; \bm{X}) - \frac{1}{2} \left[ \tr( \bm M_{J}') - \left\| \bm{x} - \bm{\mu}_{B} \right\|_{\bm M_{J}}^2 \right]
    \end{align}
    where
    \begin{align}
        \bm{M}_{J} &=  \Lambda_{B} - \Lambda_{B} \Lambda_{\tilde J}^{-1} \Lambda_{B} \\
        \bm{M}'_{J} &= \Delta_{J} \Lambda_{\tilde J}^{-1} \\ 
        \Lambda_{\tilde J} &= \Lambda_{B} + \Delta_{J}
    \end{align}
\end{proposition}
Since $\Lambda_{\tilde J} > \Lambda_B$, we have that $\bm{M}_{J} > \bm{0}$, i.e. is positive definite.
Thus, the term $\left\| \bm{x} - \bm{\mu}_{B}\right\|_{\bm M_{J}}^2 \geq 0$ for all $\bm{x}$, and $\mathbb{E} \left\| \bm{x} - \bm{\mu}_{B}\right\|_{\bm M_{J}}^2 = \tr(\bm{M}'_{J})$.
Therefore $\Swb_{\bm{Z}_{J}}(\bm{x})$ ($\Imin$) reduces to mutual information when there is only one source in $\alpha$, i.e. the self-redundancy axiom (SR).

What distinguishes $\Swb$ and PID from previous information-theoretic approaches to redundancy (e.g. interaction information~\cite{mcgill1954multivariate}) is the pointwise decomposition of information.
As we mentioned above, for the $\Imin$ PID, this pointwise decomposition is only with respect to the target variable, not the predictors.
We see this pointwise nature explicitly in the quadratic form in (\ref{eq:prop.specificInfo}).
For any realized error $\bm{x} - \bm{\mu}_B$ between the true state and the prior mean, the pointwise minimal source $\bm{Z}_{J^\star}(\bm{x})$, where $J^\star=\argmin_{J \in \alpha} \Swb_{\bm{Z}_J}(\bm{x})$, will be that source which least corrects this error --- assuming all predictor sources $J \in \alpha$ contribute information matrices $\Delta_J$ with similar profiles.

Proposition~\ref{prop.specificInfo} will follow readily from the following computation.
\begin{lemma}
\label{lemma:specificInfo.expectation_chaos_terms}
\begin{align}
\nonumber \mathbb{E} \left( \left\| 
\bm{x} - \bm{\mu}_{\tilde J} \right\|_{\Lambda_{\tilde J}}^2 - \left\| \bm{x} - \bm{\mu}_{B} \right\|_{\Lambda_{B}}^2 \; \bigg| \; \bm{x} \right) \\
= \tr(\bm{M}_{J}') - \left\| \bm{x} - \bm \mu_{B} \right\|_{\bm{M}_{J}}^2 
\label{eq:lemma.specificInfo.expectation_chaos_terms}
\end{align}
\end{lemma}

\begin{proof}
    From (\ref{eq:muTildeJ.secondMoment.general}) from Prop.~\ref{prop:muTildeJ.moments}, we compute the moment
    \begin{align}
        \nonumber \mathbb{E} \left( \left\| \bm{x} - \bm{\mu}_{\tilde J} \right\|_{\Lambda_{\tilde J}}^2  \; \bigg| \; \bm{x} \right)
        &= \tr(\Lambda_{\tilde J} \Lambda_{\tilde J}^{-1} \Delta_J \Lambda_{\tilde J}^{-1})  \\
        \nonumber &+ \left\| \Lambda_{\tilde J}^{-1}(\Lambda_B \bm{\mu}_B + \Delta_J \bm{x}) - \bm{x} \right\|^2_{\Lambda_{\tilde J}} \\
        \nonumber &= \tr(\bm{M}'_J) 
        + \left\| \bm{x} - \bm{\mu}_B \right\|^2_{\Lambda_B \Lambda_{\tilde J}^{-1} \Lambda_B}
    \end{align}
    By recombining with the conditional constant $\left\| \bm{x} - \bm{\mu}_{B} \right\|_{\Lambda_{B}}^2$, we arrive at (\ref{eq:lemma.specificInfo.expectation_chaos_terms}).
     
\end{proof}
We may now prove Proposition~\ref{prop.specificInfo}.
\begin{proof}[Proof of Proposition~\ref{prop.specificInfo}]
Using the expressions from Sec.~\ref{subsection:LFGs}, we may write the pointwise mutual information term in (\ref{eq:defn.specificQuality.PID}):
\begin{gather*}
\log \frac{p(\bm{x} | \bm{z}_{J})}{p(\bm{x})}
= \log \frac{\Phi_{\mathcal{F}_{B \cup J}}(\bm{x})}{\Phi_{\mathcal{F}_{B}}(\bm{x})} \\
= \frac{1}{2} \log \frac{|\Lambda_{\tilde J}|}{|\Lambda_B|} - \frac{1}{2} \left[ \left\| \xx - \bm{\mu}_{\tilde J} \right\|_{\Lambda_{\tilde J}}^2 - \left\| \xx - \bm{\mu}_B \right\|_{\Lambda_{B}}^2 \right]
\end{gather*}
We take this in expectation with respect to $p(\bm{z}_{J} | \bm{x})$. The first term is constant, and we use Lemma~\ref{lemma:specificInfo.expectation_chaos_terms} to integrate the second, arriving at (\ref{eq:prop.specificInfo}).
\end{proof}

\subsection{A Statistical Distance Redundancy Measure}

For our second redundancy metric, we consider a different approach.
We instead define redundancy using the statistical distances between the source-defined posterior distributions $p(\bm{X}'| \bm{Z}_J)$ and ground-truth target $\delta_{\bm{x}}$.
Our redundancy metric will look at the pointwise reduction in this statistical distance relative to that of the prior distribution $p(\bm{X}')$ from the base graph.

\begin{definition}
\label{defn:specificWER}
Let the supplemented factor graph $(\bm{X}, \mathcal{F}, B)$ be given.
We define the specific Wasserstein error reduction (WER) function to be
\begin{equation}
\label{eq:defn.specificWER}
    \Sw_{\bm{Z}_J}(\bm x) = \mathbb{E} \left( d_{\mathcal{W}^2}^2(p(\bm{X}'), \delta_{\bm{x}}) - d_{\mathcal{W}^2}^2(p(\bm{X}' | \bm{Z}_J), \delta_{\bm{x}}) \; \bigg| \; \bm{x} \right)
\end{equation}
where $d^2_{\mathcal{W}^2}$ is the square Wasserstein-2 distance. For arbitrary Gaussian density $f(\bm{u}) \sim \mathcal{N}(\bm{m},\Sigma)$ and point mass $\delta_{\bm{u}'}$, this takes the form:
\begin{equation}
\label{eq:defn.specificWER.Wass2}
    d_{\mathcal{W}^2}^2( f(\bm{u}), \delta_{\bm{u}'} ) 
    =  \tr(\Sigma) + \left\| \bm{m} - \bm{u}' \right\|^2
\end{equation}
\end{definition}

Equivalently, one could also define $\Sw_{\bm{Z}_J}(\bm x)$ as a reduction in the target-centered $L^2$-norm of the posterior pose-graph distribution, relative to the prior: 
 \begin{equation}
    \Sw_{\bm{Z}_J}(\bm x) = \mathbb{E} \left( \left\| 
 \bm{X}' - \bm{x} \right\|^2 - \left\| 
 \bm{X}'' - \bm{x} \right\|^2 \; \bigg| \; \bm{x} \right)
\end{equation}
where $\bm{X}' \sim p_{\bm{X}}$ and $\bm{X}'' \sim p_{\bm{X} | \bm{Z}_{J}}$ (\ref{eq:pxPrior}-\ref{eq:pxPosterior}).

Before we proceed to computing a closed-form for $\Sw$, we will first compute the closed form of the quality function $\Qw$ induced by Def.~\ref{defn:specificWER} through Def.~\ref{defn:targetIntegral}.
This quality function will serve a role analogous to single-source mutual information in PID.
\begin{proposition}
    The quality function for the redundancy measure induced by Def.~\ref{defn:specificWER} has the closed form:
    \begin{equation}
        \Qw(\bm{Z}_J) = 2 \tr(\Lambda_{B}^{-1} - \Lambda_{\tilde J}^{-1})
    \end{equation}
    Moreover, this function satisfies monotonicity (MQ).
\end{proposition}
\begin{proof}
    Combining Defs.~\ref{defn:targetIntegral}\&\ref{defn:specificWER}, we have that 
    \begin{align}
        \nonumber \Qw(\bm{Z}_{J}) &= \mathbb{E} d_{\mathcal{W}^2}^2(p(\bm{X}'), \delta_{\bm{X}})\\
        &- \mathbb{E} d_{\mathcal{W}^2}^2(p(\bm{X}' | \bm{Z}_J), \delta_{\bm{X}})
    \end{align}
    Since $\mathbb{E} \left| \bm{\mu}_{J'} - x \right\|^2 = \tr(\Lambda_{J'}^{-1})$ for all $\bm{Z}_{J'} \subset \bm{Z}$, we use (\ref{eq:defn.specificWER.Wass2}) to find that find that $\mathbb{E} \, d_{\mathcal{W}^2}^2(p_{\bm{X}}, \delta_{\bm{X}}) = 2 \tr(\Lambda_B^{-1})$ and $\mathbb{E} \, d_{\mathcal{W}^2}^2(p_{\bm{X}|\bm{Z}_J}, \delta_{\bm{X}}) = 2 \tr(\Lambda_{\tilde J}^{-1})$.
    The monotonicity of $\mathtt{Q}$ follows from Prop.~\ref{prop:Loewner.subsource} and the monotonicity of the trace.
\end{proof}

We now compute the closed-form solution of our specific quality function.
\begin{proposition}
\label{prop.specificWER}
    \begin{align}
        \nonumber \Sw_{\bm{Z}_J}(\bm x)
        &= \tr \left( \bm{N}'_J \right) 
        + \left\| \bm{\mu}_B - \bm{x} \right\|^2_{\bm{N}_J}
 \label{eq:prop.specificWER} \\
 \intertext{where}
     \bm{N}'_J
     &= \Lambda_B^{-1} - \Lambda_{\tilde J}^{-1} 
     - \Lambda_{\tilde J}^{-1} \Delta_{\tilde J} \Lambda_{\tilde J}^{-1} \\
     \bm{N}_J
     &= I - \Lambda_B \Lambda_{\tilde J}^{-2} \Lambda_B
    \end{align}
\end{proposition}
\begin{proof}
    From (\ref{eq:defn.specificWER}) and (\ref{eq:defn.specificWER.Wass2}), we have that
    \begin{align}
        \nonumber \Sw_{\bm{Z}_J}(\bm x)
        &= \tr \left(\Lambda_B^{-1} - \Lambda_{\tilde J}^{-1} \right)\\
        &+ \mathbb{E} \left(
        \left\| \bm{\mu}_B - \bm{x} \right\|^2 - \left\| \bm{\mu}_{\tilde J} - \bm{x} \right\|^2  \; \bigg| \;  \bm{x} \right)
        \label{eq:proof.specificWER.expandDiff}
    \end{align}
    As a special case of the moment (\ref{eq:muTildeJ.secondMoment.general}) from Prop.~\ref{prop:muTildeJ.moments}, we have
    \begin{align}
        \nonumber \mathbb{E} \left( \left\| \bm{\mu}_{\tilde J} - \bm{x} \right\|^2 \; \bigg| \;  \bm{x} \right)
    &= \tr \left(\Lambda_{\tilde J}^{-1} \Delta_J \Lambda_{\tilde J}^{-1} \right) \\
    \label{eq:muTildeJ.secondMoment.recentered}
    &+ \left\| \bm{\mu}_B - \bm{x} \right\|^2_{ \Lambda_B \Lambda^{-2}_{\tilde J} \Lambda_B }
    \end{align}
    Substituting this into (\ref{eq:proof.specificWER.expandDiff}) and recombining terms, we arrive at (\ref{eq:prop.specificWER}).
\end{proof}

\section{An Application to 2D Landmark SLAM}
\label{sec:SLAM_application}
In order to investigate the plausibility and behavior of our proposed $\Rwb$ and $\Rw$ redundancy metrics, we now apply them to simulations of 2D landmark-based SLAM task.
In our simulation, a single robot randomly traverses a 2D environment populated by two landmarks.
By taking measurements of both landmarks, it will supplement its pose graph with additional, potentially redundant information.
We will use our redundancy metrics to estimate the redundancy that comes from using both landmarks, rather than one.
In these simulations, we dispense with the linearity restriction we have assumed up this point.
Our robot's configuration space will be the 2D rigid body transformations $\SE(2)$, representing its pose (translation and rotation) relative to the origin in $\mathbb{R}^2$.
For all factor graphs operations, we use the GTSAM~\cite{gtsam}  toolbox for nonlinear factor graphs.
Allow $y^{\text{t}} \in \mathbb{R}^2$ to represent the translation of any $y \in \SE(2)$, and $\bm{y}^{\text{t}} = (y_i^{\text{t}})_{i=1}^n$ for any vector $\bm{y} \in \left( \SE(2) \right)^n$.

We denote the robot's pose trajectory as
$\bm{X} = (X_0, ..., X_n)$, taking values in $SE(2)$ over $n$ timesteps, and let $L_0$ and $L_1$ denote the positions of the landmarks in $\mathbb{R}^2$. Collectively,
these $n+2$ variables are represented in the robot's factor graph, yet we will compute the redundancy metrics only with respect to trajectory variables $\bm{X}$.
Both the robot's initial position and those of the landmarks are independently sampled from uniform distributions at initialization,
i.e $L_0, L_1 \sim \mathcal{U}([-C,C]^2)$ and $X_0^{\text{t}} \sim \mathcal{U}([-C/2,C/2]^2)$ for some $C>0$.
We generate the robot's trajectory with a stationary random walk, i.e. $\eta_i = X_{i-1}^{-1} X_i$ is sampled with the same mean and covariance for each $i$.
At each step $i=1,...,n$, the robot takes an unbiased odometry measurement $Z_i$ of $\eta_i$ with fixed noise matrix $\Sigma_{\text{odom}}$ in $(x,y,\theta)$-coordinates.
These odometry measurements compose the base graph.
Moreover, at every timestep $i$, the robot takes a range-bearing measurement of each landmark, indexed $Z_{(s+1)n+i}$ for landmark $s \in \{0,1\}$, and adds it to its factor graph.
The bearing measurement has fixed variance, while the variance of the range measurement scales quadratically with distance.

We will be estimating the redundancy of the 2-antichain $\alpha = \{ \bm{Z}_{J_0}, \bm{Z}_{J_1} \}$ where $J_s = \{ i | (s+1)n < i \leq (s+2) n\}$ --- that is, the redundancy between the cumulative measurements of $L_0$ and those of $L_1$.
In order to compute $\Rwb$ and $\Rw$, we compute the integral in (\ref{eq:defn.targetIntegral.redundancy}) by sampling from the base graph $\bm{X}' \sim p_{\bm{X}}$.
We linearize each subgraph $\mathcal{F}_{\tilde J_s}$ about its solution $\bm{\mu}_{\tilde J}$.
We then estimate each of the specific quality functions $\mathtt{S}_{\bm{Z}_{\tilde{J}_s}}$ at $\bm{x}$ using the linear formulae from Props.~\ref{prop.specificInfo} and \ref{prop.specificWER} as a first approximation.

\begin{figure}[ht!]
    \centering
    \vspace{-0.5cm}
    \includegraphics[scale=0.57]{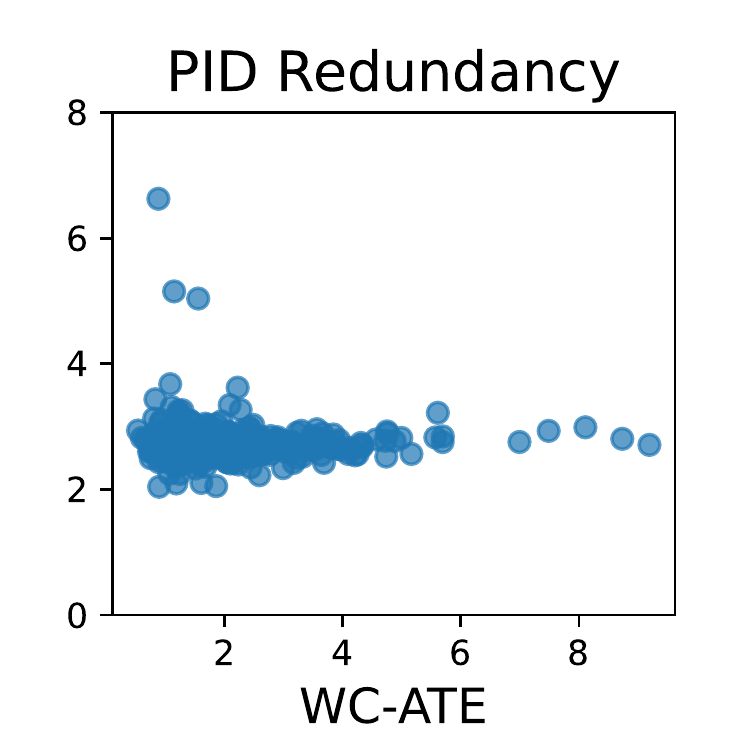}
    \includegraphics[scale=0.57]{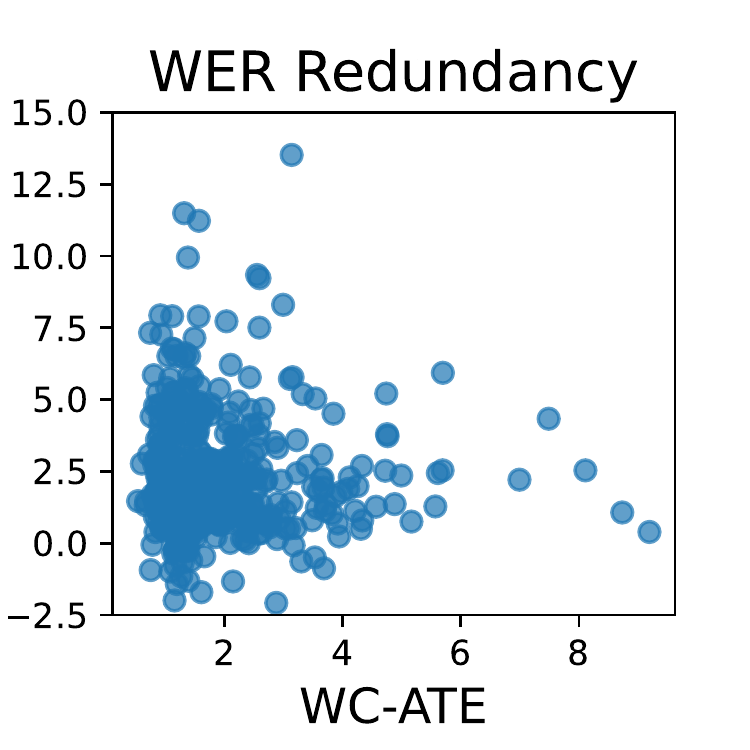}
    \vspace{-0.7cm}
    \caption{\textbf{Redundancy and the worst-case ATE.} For 500 simulations, we present the relationship between the estimated $\Rwb$ and $\Rw$ redundancies and WC-ATE, as computed from ground-truth. As can be seen, for the $\Rw$ redundancy metric, factor graphs with less accurate single-landmark subgraphs (higher WC-ATE) tend to have low estimates of redundancy between the two landmarks. Additionally, simulated factor graphs with high redundancy estimates tend to have low WC-ATE. However, the converse of these statements is not true: many simulations had low redundancy estimates and low WC-ATE. }
    \label{fig:landmarkSLAM.results.WCATE}
    \vspace{-0.1cm}
\end{figure}

Given ground-truth, one can evaluate the localization component of a SLAM pipeline using absolute or relative trajectory error (ATE and RTE, respectively) \cite{zhang2018tutorial}.
It is less immediately obvious what ground-truth redundancy estimates should be compared against.
We will define a variant of ATE that aims at ground-truth error from a redundancy perspective.

\begin{figure*}[ht!]
\centering
\includegraphics[scale=0.48]{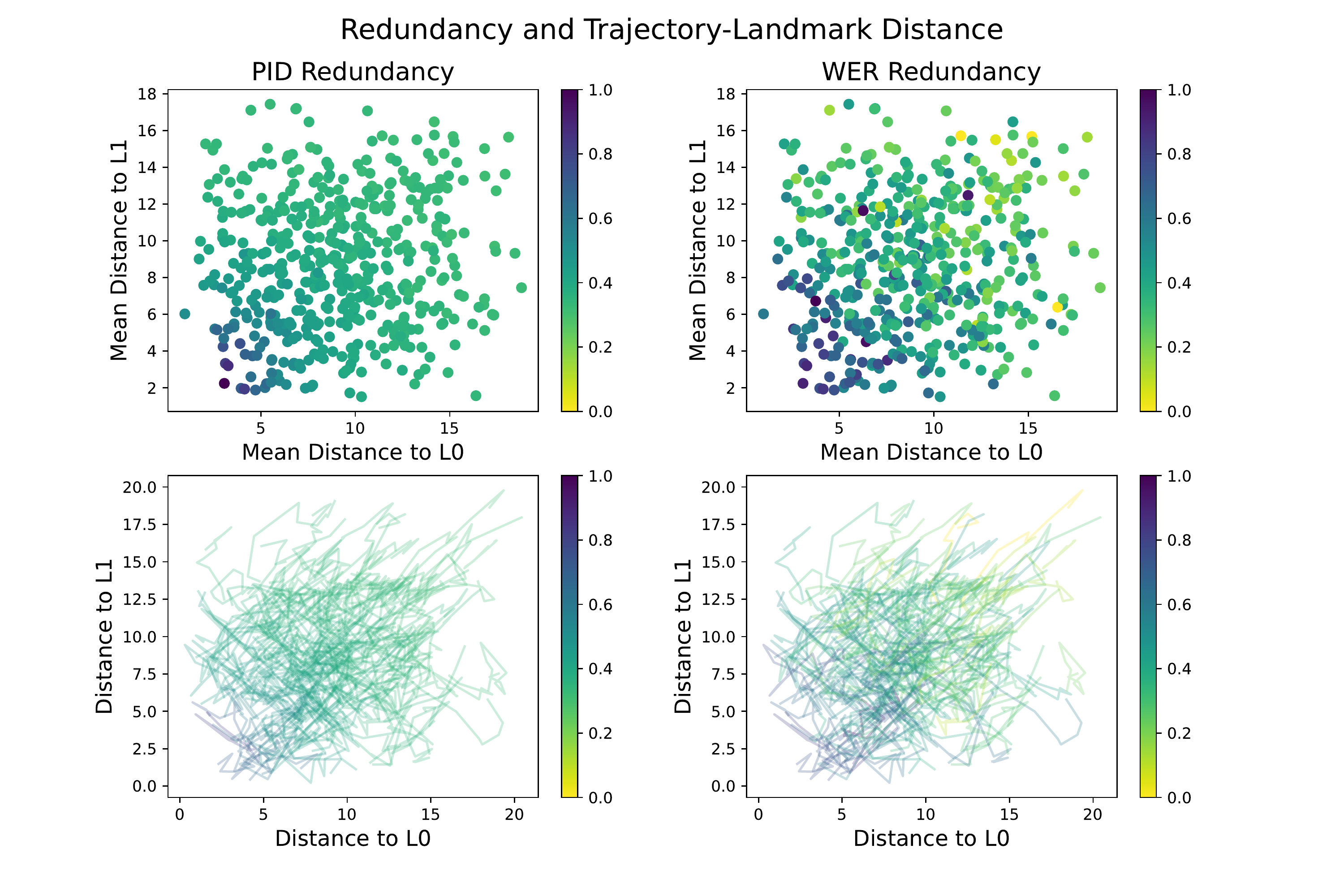}
\vspace{-0.7cm}
\caption{\textbf{Redundancy and landmark distance.} We visualize the relationship between our redundancies and the distance of the robot's trajectory from the landmarks.
(TOP) We represent the relationship between the mean distances of the trajectory to each landmark, and the batch-normalized redundancy scores.
(BOTTOM) In order to better account for the heterogeneity of posewise distances along a trajectory, we visualize the landmark distances posewise for the first 200 simulations, with each line connecting the points corresponding to the 10 poses of a given simulation.
As can be seen, trajectories that pass or remain close to both landmarks are characterized by the highest redundancy scores, while drifting away from either landmark reduces redundancy.
}
\label{fig:landmarkSLAM.trajDist}
\vspace{-0.4cm}
\end{figure*}

\begin{definition}[Worst-Case Absolute Trajectory Error]
    Let the nonlinear supplemented factor graph $(\bm{X}, \mathcal{F}, B)$ be given for ground-truth trajectory $\bm{x} \in \SE2^{n}$.
    Moreover, let an antichain of supplemental factors $\alpha = (\bm{Z}_{J_1}, ... \bm{Z}_{J_k})$ be given as well, with $\bm \mu_{\tilde J_k}$ as the MAP estimate of $\bm{X}$ from each $\mathcal{F}_{\tilde J_k}$.
    The worst-case absolute trajectory error (WC-ATE) is given by:
    \begin{equation}
        \text{WC-ATE}(\bm{x}, \bm{z}) = \sum \max_{J_k} \left\| \bm{x}^{\text{t}}_i - \bm{R}_{J_k} \mu_{\tilde J, i}^{\text{t}} - t_{J_k} \right\|^2
    \end{equation}
    where $\bm{R}_{J_k}, t_{J_k} = (\bm{R},t)(\bm{x}^{\text{t}},\bm \mu_{\tilde J_k}^{\text{t}})$ are the transformation and translation from the Umeyama method for aligning point-sets \cite{umeyama1991least}.
\end{definition}
In this case, we compute the WC-ATE for the antichain $\alpha = \{ \bm{Z}_{J_0}, \bm{Z}_{J_1} \}$ to compare to the redundancy scores.

We present exploratory results from our experiments in Figs.~\ref{fig:landmarkSLAM.results.WCATE}-\ref{fig:landmarkSLAM.trajDist}.
In Fig.~\ref{fig:landmarkSLAM.results.WCATE}, we plot the WC-ATE of $500$ simulated trajectories of $n=10$ poses against the $\Rwb$ and $\Rw$ redundancies estimated from the robot's factor graph.
For the $\Rw$, we see that higher redundancies suggest, with high probability, a lower WC-ATE --- i.e, the error in any pose is likely to be small regardless of which landmark is used for SLAM.
However, the converse is not true: lower estimated $\Rw$ does not necessarily imply that both landmarks are needed in every realization.
This is expected: WC-ATE is based upon the fully realized trajectory $\bm{x}$ and measurements $\bm{z}$, whereas the redundancies are computed in expectation.
Put differently, it will often be the case that $\bm{\mu}_{\tilde J_s}$ for both $s=0,1$ will be closer to $\bm{x}$ in realization than expectation.

In Fig.~\ref{fig:landmarkSLAM.trajDist}, we explore a possible source of heterogeneity between simulations that we would expect to affect redundancy: the distance of the landmarks to the robot.
Since the simulated range-bearing measurements degrade with distance, we expect the information quality contributed by each landmark to scale down with distance.
Moreover, we expect maximal redundancy in the contributed quality when both landmarks are close to the robot.
We see that this is generally the case: the highest redundancy scores for both $\Sw$ and $\Swb$ were computed from trajectories that remained close to both landmarks.

\section{Conclusion and Future Work}

In this paper, we have begun the exploration in the formal and theoretical direction for developing a theory of multi-source redundancy for factor graphs, drawing upon ideas from the partial information literature.
We adapt the PID axiomatic framework to the factor graph setting, generalizing the language of redundancy for quality metrics beyond mutual information (as in PID).
Within this framework, we contribute two redundancy metrics on linear factor graphs.
These metrics were applied to simulations of a 2D SLAM scenario.

We are currently working toward creating a rigorous toolbox for estimating and employing redundancy in linear factor graphs.
In future work, we will generalize the redundancy metrics proposed to nonlinear factor graphs.

\bibliographystyle{ieeetr}
\bibliography{references}

\vspace{12pt}

\end{document}

%% file: preamble.tex
\usepackage{amsmath,amsfonts,amssymb,amsthm}
\usepackage[utf8]{inputenc}
\usepackage{verbatim}
\usepackage{graphicx}
\usepackage{amsfonts}
\usepackage{soul}

\usepackage{amsmath,amsfonts,amssymb}
\usepackage{graphicx}
\usepackage[colorlinks=true, allcolors=blue]{hyperref}

\usepackage{bm}

\newtheorem{corollary}{Corollary}
\newtheorem{lemma}{Lemma}

\newtheorem{proposition}{Proposition}
\newtheorem{definition}{Definition}

\newtheorem*{remark}{Remark}

\newcommand{\SE}{\text{SE}}

\newcommand{\xx}{\bm{x}}

\newcommand{\bb}{\bm{b}}

\DeclareMathOperator{\rank}{rank}

\newcommand{\MIN}{\text{min}}
\newcommand{\Wass}{\mathtt{W}_2}

\newcommand{\WB}{\mathtt{WB}}

\newcommand{\Ired}{I_{\cap}}
\newcommand{\Imin}{I_{\cap}^{\MIN}}

\newcommand{\Rw}{\mathtt{R}^{\Wass}}
\newcommand{\Rwb}{\mathtt{R}^{\WB}}
\newcommand{\Sw}{\mathtt{S}^{\Wass}}
\newcommand{\Swb}{\mathtt{S}^{\WB}}
\newcommand{\Qw}{\mathtt{Q}^{\Wass}}
\newcommand{\Qwb}{\mathtt{Q}^{\WB}}

\usepackage{tikz}
\usetikzlibrary{shapes.geometric, arrows,decorations.pathreplacing}
\tikzstyle{pose} = [circle, text centered, draw=black, fill=orange!30, minimum size = 1.25cm]
\tikzstyle{spur} = [circle, text centered, draw=black, fill=violet!30, minimum size = 1.25cm]
\tikzstyle{factor} = [rectangle, text centered, draw=black, fill=blue!30]
\tikzstyle{rendezFactor} = [rectangle, text centered, draw=black, fill=violet!30]
\tikzstyle{rendezFactorEliminated} = [rectangle, text centered, draw=black!40, text = black!40, fill=violet!10]
\tikzstyle{factorNew} = [rectangle, text centered, draw=black, fill=green!30]
\tikzstyle{factorNewEliminated} = [rectangle, text centered, draw=black!40, text=black!40, fill=green!10]

\tikzstyle{poseEliminated} = [circle, text centered, draw=black!40, text=black!40, fill=orange!10]
\tikzstyle{spurEliminated} = [circle, text centered, draw=black!40, text=black!40, fill=violet!10, minimum size = 1.25cm]
\tikzstyle{factorEliminated} = [rectangle, text centered, draw=black!40, text=black!40, fill=blue!10]

\tikzstyle{intra} = = [thick,->,>=stealth]
\tikzstyle{inter} = = [dashed,-, very thick,draw=blue]
\tikzstyle{factorLine} = = [thick,-]
\tikzstyle{factorLineEliminated} = = [thick,-,draw=black!40]
\tikzstyle{factorLineNew} == [thick, green, -]
\tikzstyle{BNarrow} = = [dashed, blue, ->]
\tikzstyle{interrobotComms} = [dashed,->,draw=violet,thick]

\DeclareMathOperator{\tr}{tr}

\DeclareMathOperator{\argmin}{argmin}

\usepackage{todonotes}
\usepackage[most]{tcolorbox}

\usepackage{dirtytalk}

\usepackage[framemethod=TikZ]{mdframed}
\newcounter{assumption}[section]
\renewcommand{\theassumption}{\arabic{section}.\arabic{assumption}}

\newcounter{metric}[section]
\renewcommand{\themetric}{\arabic{section}.\arabic{metric}}
